\documentclass[5p]{elsarticle}

\usepackage{lineno}
\usepackage[bf]{subfigure}
\usepackage{color}

\switchlinenumbers
\modulolinenumbers[5]

\biboptions{authoryear}

\definecolor{revisioncolor}{rgb}{0.12,0.47,0.71}

\begin{document}

\begin{frontmatter}

\title{Using network science and text analytics to produce surveys in a scientific topic}

\author[ifsc]{Filipi N. Silva}

\author[icmc]{Diego R. Amancio\corref{mycorrespondingauthor}}
\ead{diego@icmc.usp.br}

\author[tyndall]{Maria Bardosova}

\author[ifsc]{Luciano da F. Costa}

\author[ifsc]{Osvaldo N. Oliveira Jr}

\address[ifsc]{S\~ao Carlos Institute of Physics, University of S\~ao Paulo,  S\~ao Carlos, Brazil}
\address[icmc]{Institute of Mathematics and Computer Science, University of S\~ao Paulo, S\~ao Carlos, Brazil}
\address[tyndall]{Tyndall National Institute, Cork City, Ireland}
\cortext[mycorrespondingauthor]{Corresponding author}

\begin{abstract}
The use of science to understand its own structure is becoming popular, but understanding the organization of knowledge areas is still limited because some patterns are only discoverable with proper computational treatment of large-scale datasets. In this paper, we introduce a framework to combine network-based methodologies and text analytics to construct the taxonomy of science fields. The methodology is illustrated with application to two topics: \emph{complex networks} (CN) and \emph{photonic crystals} (PC). We built citation networks using data from the Web of Science and used a community detection algorithm for partitioning to obtain \emph{science maps} for the two topics. We also created an importance index for text analytics, which is employed to extract keywords that define the communities and, combined with network topology metrics, to generate dendrograms of relatedness among subtopics. Interesting patterns emerging from the analysis included identification of two well-defined communities in PC area, which is consistent with the known existence of two distinct communities of researchers in the area: telecommunication engineers and physicists. With the methodology, it was also possible to assess the interdisciplinary nature and time evolution of subtopics defined by the keywords. The automatic tools described here are potentially useful not only to provide an overview of scientific areas but also to assist scientists in performing systematic research on a specific topic.
\end{abstract}


\end{frontmatter}



\section{Introduction}
Recent developments in the use of machine learning methods to extract information (and knowledge!) from Big Data have shown that machines are bound to replace humans in various intellectual tasks in the near future, particularly in cases where a lot of information needs to be processed~\citep{Craddock:2008eS,Donovan:2008aa,Bell06032009}. Clear examples of such tasks are facial recognition~\citep{zhao2003face}, establishing best routes for cars and passengers~\citep{laporte1992thevehicle}, internet search~\citep{lawrence1998searching}, etc. Some authors have even been bold enough to suggest that scientific and technological development is being held back by the limited capacity of humans, especially the memory, to process and interpret the electronic data available~\citep{Stone:2014aa}. A specific task in academic work where this limited capacity is readily apparent is in carrying out a survey of any given topic, owing to the vast literature to be consulted. The first requirement for a survey, namely to establish a map of knowledge (also known as \emph{science map}) of the field under analysis, demands data-intensive discovery. Surveys normally performed by humans benefit from well-founded techniques to organize scientific literature and information, but little help exists for understanding the knowledge structure on a larger scale. Even experienced researchers find this hard owing to the aforementioned human limited capacity, and there is the additional drawback of bias - even if unintentional - toward the experts' personal preferences. Not surprisingly, modeling the knowledge structure remains an open problem in science with the intricate relationships among the many concepts involved.

In this paper we propose a new framework to assist humans in preparing literature surveys, which consists of the integration of many well-established concepts arising from complex networks~\citep{Barabasi:1999p279} that have been proven effective in modeling the organization of knowledge~\citep{Boyack:2005fk,Borner:2009aa,Costa2011analyzing, silva:2013, boyack2014Creation}. Our approach, however, distinguishes itself from previous ones in the literature since network science and text analytics methods are interwoven to generate science maps and taxonomies. More specifically, we build citation networks~\citep{chen2004tracing,menczer2004correlated,Leicht:2007aa} that serve as the overall framework of a science map, which needs to be complemented with a taxonomy to classify the contents of the map. We adapted the methodologies to extract keywords to complete the science map for two fields, namely ``Complex Networks'' and ``Photonic Crystals''. This choice was basically due to the authors of the paper being experts in these fields, which allows for a deeper discussion of the results obtained.

\section{Overview of Complex Networks and Text analytics applied to summarization}
Because our study deals with two very distinct areas, namely use of complex network methods to analyze scientific literature and text analytics, a brief overview of previous work will be done here for these areas. This overview is by no means exhaustive, particularly as there has been a vast literature in each of these areas; we rather concentrate on work that is directly related to the purpose of our study, which is to provide semi-automated means for assisting authors in surveys of the literature and document summarization techniques~\cite{silva:2011}.

Recent works have used network-based metrics to characterize or quantify relevance and impact of researchers, publications and journals~\citep{Ding:2009aa,Yan:2013aa,Nykl:2015aa,Zhou:2015aa,McKeown:2016aa}. For instance, factor analysis was employed to automatically extract the most important papers in citation networks~\citep{TsungTeng:2012}. Citation-based networks have been used in various domains, such as modeling the dynamics of knowledge acquisition and dissemination~\citep{Borner:2009aa,Amancio:2012aa,txiart}, enriching and contextualizing information of biological experiments or data~\citep{Mullen:2014aa}, and visualizing relationships among scientific fields by constructing science maps~\citep{Boyack:2005fk, Leydesdorff:2009aa,Boerner:2012aa}.

Of particular importance are science maps used as a versatile tool to qualitatively understand how science fields are organized, by e.g. establishing relationships among distinct areas~\citep{Boyack:2005fk, Leydesdorff:2009aa, Porter:2009aa, rosvall2008maps, Silva:2010aa,silva:2011}. Tools have been developed to visualize and interact with scientific maps~\citep{Boyack:2002aa,Eck:2010aa,Waaijer:2011aa,network3D:2015,silva:2013,Eck:2014aa}, and understand interdisciplinarity~\citep{Porter:2009aa,Leydesdorff:2013aa, silva:2013, Lariviere:2015aa, Leydesdorff:2015aa} among scientific journals. In a similar fashion, science maps can also be constructed by using self-organizing maps in which scientific domains are mapped to a 2D space according to a neural network through a Hebbian learning process~\citep{Skupin:2013aa}. While science maps are able to provide interesting insights about the overall structure of science, a contextualized taxonomy of its structure is more appropriate to the task of surveying a scientific field. This is because survey papers are conventionally organized in a hierarchical structure, normally comprising chapters, sections, subsections and other forms of text partitions. Establishing such taxonomy, with components and subcomponents hierarchically organized, is not trivial for automated tools~\citep{Sebastiani:2002,silva:2013}, and various procedures have been adopted to classify contents.

Text summarization is a traditional area of text analytics, which has been used to build summaries and taxonomies of text datasets comprising many types of situations, such as tracing the events of disasters using social media~\citep{McKeown:2015}, conferences~\citep{Shen:2013aa} and sports events~\citep{Nichols:2012aa}. The main goal with such techniques is to obtain an importance metric (also called \emph{salience}) for terms or sentences. The summary of the content can be constructed by rewriting the text using only terms or sentences presenting high salience, while the taxonomies can be obtained by clustering texts according to the similarities among their most important terms. This can be accomplished through the use of metrics such as cosine similarity~\citep{Salton:1988aa} or semantic-wise similarities~\citep{Boyack:2011aa}, as in relationships in the WordNet or word embedding techniques~\citep{Levy:2014aa}. A simple way to obtain the salience of terms is by comparing their relative frequency of appearance inside a document to their frequency of appearance in a larger set of other documents. This is usually referred to as the TF-IDF~\citep{Salton:1988aa} method, which yields good results for sets of large texts. However, the method becomes unreliable when measuring relevance of terms in sets of small texts, since terms tend to appear only a few times for each document, as in paper abstracts and messages of social networking services. Other, more complex, summarization techniques can be used to deal with such type of data. Examples are supervised machine learning methods that require a small set of golden summaries used to train a machine to detect important terms. Human readable summaries may be generated from a document or a set of documents~\citep{Radev:1998aa} by using features of low contextual content, such as the average number of words or the  number of capitalized words in a sentence~\citep{McKeown2012}.

As an alternative to machine learning methods, topics analysis~\citep{Blei:2003aa} has been employed to find important terms (keywords) in a set of documents, such as articles or abstracts~\citep{Griffiths:2004aa}, where terms are projected and clustered according to their presence in a set of documents. This is done by estimating a Markov chain model of topic information along the documents, normally obtained by Gibbs sampling. This technique presents high computational cost, as it requires several iterations to estimate the transitions between words, but it can give good results depending on the size of each document, the number of documents and other properties of the dataset, as studied in depth by Tang \emph{et. al} in Ref.~\citep{icml2014c1_tang14}.

Methods derived from network science have also been used for document summarization. The LexRank technique~\citep{Erkan:2004aa} relies upon a network of similarity between sentences to obtain topological centrality measurements, such as eigenvector centrality~\cite{Newman:2010:NI:1809753}. The centrality measurements are then used to quantify the salience of terms. In a similar fashion, word adjacency networks were employed to find keywords in a text~\citep{Amancio:2012ab}, where salience was obtained from the diversity measurement of nodes~\citep{Viana:2010aa} and provided superior results to traditional centrality measurements in networks. Such kind of analysis is advantageous compared to multiple text analytics methods for the same dataset since information provided by network based techniques does not overlap with that provided by traditional text analytics techniques~\citep{Li20125254,DBLP:journals/corr/NewmanC15,Amancio:2015aa,0295-5075-98-5-58001,0295-5075-99-4-48002}.

\section{Methodology}

\begin{figure}[!htbp]
 \centering
 \includegraphics[width=0.9\linewidth]{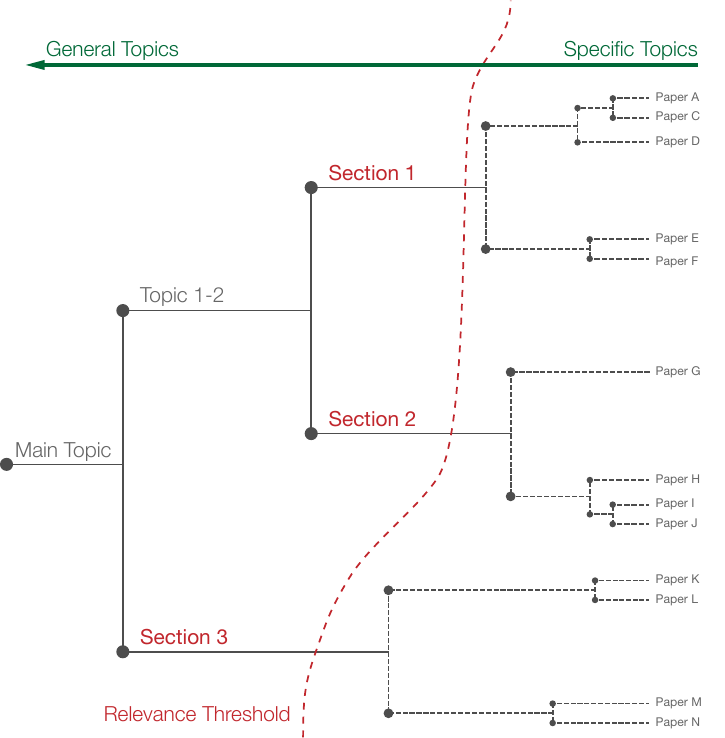}
 \caption{Example of the structure of an organized scientific survey. Papers are grouped into more general topics which are reflected as sections, subsections, chapters, etc. A threshold of relevance and focus is thus necessary as their content needs to be summarized and cannot retain the full level of detail for each paper.}
 \label{fig:dendrogramExample}
\end{figure}

A survey paper is taken here as an organized structure that summarizes information about a scientific field. It must limit the level of detail for each topic by highlighting the most relevant pieces of information while also reducing their redundancy. The hierarchy in a survey comprises concepts that are progressively merged together by their relatedness to build major contextual structures such as subsections and sections, as exemplified in Fig.~\ref{fig:dendrogramExample}. Topics are hierarchically structured, each of which can represent a set of papers or other scientific works relevant to the area.

To determine the hierarchy of components and subcomponents, we first built citation networks for the fields \emph{Complex Networks} (CN) and \emph{Photonic Crystals} (PC), whose papers were retrieved from the Web of Science (WOS)\footnote{\url{http://thomsonreuters.com/thomson-reuters-web-of-science/}} database using the query terms "complex network" and "photonic crystal" (including the plural variations), respectively. For each retrieved paper, we extracted the title, abstract, publication year, citation count and list of references. Two citation networks were built (CN and PC) where nodes represent the papers and an edge was established between two papers if one cites the other.

There are many ways to construct citations-based networks. They can be drawn directly from the citation structure, in which two papers are connected if there is a citation between them, resulting in an unweighted directed network. Also used in several studies are co-citation networks~\citep{Usdiken:1995aa,Jenssen:2001aa,Chen:2004aa,Ding:2009aa}, where documents are connected if they share a citation with at least another document. This procedure leads to a weighted undirected network, and the number of shared documents can be used as a metric of similarity among documents.

For the sake of simplicity, here we opted to use traditional citation networks, but we do not take into account the direction of citation connections. We understand that this information is relevant in several other studies~\citep{chen2004tracing,menczer2004correlated}, but not here because we use citation networks to represent a knowledge relationship structure which is naturally undirected. As an alternative, we also applied the analysis presented in this work to co-citation networks as shown in the supplementary material, and found similar results in the analysis. However, such networks are denser and harder to discuss and visualize.

The citation networks were constructed by first obtaining the vertices from papers returned from the chosen queries for CN and PC in the Web of Science dataset. Next, citation information was used to connect pairs of cited papers where papers that were not present in the initial queries were ignored (even if cited by others). This avoids problems caused by dangling nodes, which can impact the topological analysis employed here, such as community detection.

Citation networks can be transformed into science maps if the most relevant topics and their inter-relationships are identified. In this study, the CN and PC citation networks were embedded in a 3D space using a force-directed method based on the Fruchterman-Reingold algorithm~\citep{FRUCHTERMAN:1991fk}. The initial configuration had the nodes, treated as particles, uniformly distributed over a 3D space. These nodes were allowed to interact via repulsive forces, with attractive forces being added for the connected nodes. When the energy of the whole system was minimized, the resulting embedding became a graphically appealing projection of the networktopology~\citep{silva:2013,Bando:2013}. In print, only static 2D projections of the network can be visualized, but the network structure can be further examined with a visualization tool~\citep{silva:2013,Bando:2013}. This is important because real system topologies may exhibit very high dimension, hence not suitable to be projected on the plane~\citep{daqing2011dimension}.

The main topics in a field are associated with communities in the citation networks, which were determined by applying the multilevel community detection method~\citep{blondel2008fast}. This procedure assigns each paper to a non-overleaping community. It was chosen because it allows for establishing a high modularity for the network, while keeping the computational cost reasonable in comparison with more sophisticated methods such as the optimum modularity~\citep{newman2006modularity}. By a high modularity we mean that the communities in the network are well distinguishable from each other. It is important to highlight that the multilevel community detection method is stochastic, thus, for each run, a distinct community structure can be attained for the same network. However, as discussed in ref.~\citep{blondel2008fast}, the resulting community partitioning for distinct runs are very similar among themselves and display high correspondence to those obtained by other algorithms or expected from benchmarks.

The relationships among communities were further examined by generating a coarse-grained graph of the network~\citep{rosvall2008maps}, in which each community was replaced by a single community node and its connections. The edges between each pair of community nodes $(\alpha,\beta)$ were weighted by $W_{\alpha \beta}$ according to the stochastic probability of connections between communities $\alpha$ and $\beta$ given by:
\begin{equation}
	W_{\alpha \beta} = {E_{\alpha \beta} \over |\alpha| |\beta|},
\label{eq:normalized_mixing}
\end{equation}
where $E_{\alpha \beta}$ is the number of connections among nodes of communities $\alpha$ and $\beta$.

Since determining the communities which are most central or peripheral in the science map is an important target, we employed the \emph{accessibility} metric~\citep{travenccolo2008accessibility,travencolo2009border,Arruda:2014aa,Amancio:2015aa}, which is a local node-centered measurement based on the heterogeneity of probabilities of reaching nodes in random walk dynamics. The smaller the accessibility of a node the more peripheral it is. This metric has been successful in separating the topological center and border regions of networks while avoiding the drawbacks of traditional measurements such as betweenness centrality.

Ideally, the communities in the citation network should be labeled with the topics and subtopics of a well-established taxonomy for the scientific field under analysis. However, as already mentioned in the Introduction, there is no simple way to generate such high-level taxonomy automatically. Most authors have therefore resorted to extracting keywords (see Refs.~\cite{Andrade01011998, Manning:1999:FSN:311445, hulth2003improved, CarreteroCampos2013} for methods of keyword extraction), for which the majority of the methods make use of large amounts of text. In our case, because we only considered the Abstracts from each paper (representing a node in the network), we had to adapt existing methods. We devised a measurement to quantify the importance of keywords, made with unigrams and bigrams, for each network community. Unigrams and bigrams were extracted for each paper by analyzing its abstract, from which stop-words were removed and the remaining words were lemmatized. This pre-processing step is essential for the analysis because it removes words conveying little semantic content and semantically related words are aliased under the same word if they share the same canonical form~\citep{0295-5075-100-5-58002,1367-2630-14-4-043029,txiart}. The \emph{importance index} was designed to quantify the relative frequency of a word appearing inside a community against its frequency on the remainder of the network. First, we count the total number of times $n_\alpha(w)$ a paper presenting a word $w$ appears inside a community $\alpha$. Next, we calculate the relative in-community frequency, $F^{in}_\alpha(w)$ given by:
\begin{equation}
F^{in}_\alpha(w) = {n_\alpha(w) \over {|\alpha|}},
\label{eq:intraFrequency}
\end{equation}
where $|\alpha|$ is the number of papers associated with a community $\alpha$. Analogously, we define a relative out-community frequency:
\begin{equation}
F^{out}_\alpha(w) = \sum_{\gamma \neq \alpha} {n_\gamma(w) \over N-|\alpha|},
\label{eq:outerFrequency}
\end{equation}
which accounts for the total relative frequency considering all communities excluding $\alpha$, where $N$ is the total number of papers in the network. Then, we define our measurement of importance of keywords, $I(w)$, as the highest difference between the relative in-community and out-community frequencies of a word:

\begin{equation}
I(w) = \max_\alpha[F^{in}_\alpha(w) - F^{out}_\alpha(w)].
\label{eq:keywordRanking}
\end{equation}

The keywords ranked according to the importance index $I(w)$ were used to create trees to simulate the structure of a survey, as shown Figure~\ref{fig:dendrogramExample}. The hierarchy tree (dendrogram) was obtained by a hierarchical agglomerative clustering method~\citep{Duda:2001qc,Costa:2009ss}, in which we used the average shortest path length, $\langle\ell\rangle_{uv}$, among pairs of keywords $(u,v)$. In this procedure, we first obtained the shortest path lengths $\ell_{ij}$ between the pairs of papers $(i,j)$ in the citation network. Next, for each keyword pair $(u,v)$ we calculated the average of $\ell_{ij}$ among pairs of abstracts $(A_i, A_j)$ of papers $(i, j)$, where the keywords $u$ and $v$ were respectively present. This can also be written by the following equation:
\begin{equation}
\langle\ell\rangle_{uv} = \sum_{(u,v)\,\in\,(A_i\times A_j)}  {\ell_{ij} \over |(u,v)\,\in\,(A_i\times A_j)|}.
\label{eq:MetricClustering}
\end{equation}
As a consequence, groups of keywords are progressively clustered together according to the average topological distance between them. Therefore, our approach to generating dendrograms incorporates both concepts from complex networks and from text analytics. This was crucial because clustering the keywords using only the Abstracts would not be precise as the amount of text is limited.

Since unigrams and bigrams were ranked according to the same measurement, if a bigram has high $I(w)$, their compounded unigrams are very likely to also feature among the top keywords. To address this problem, we removed the unigrams from the set of keywords that are part of any other bigram in the set. By doing this, we eliminate an immediate layer of redundancy among keywords while also giving priority to more specific keywords(bigrams). For the PC field, we also generated a dendogram using keywords suggested by an expert, in which we omitted generic keywords covering more than $50\%$ of the network (e.g. \emph{photonic crystals} and \emph{fiber}).

The temporal evolution of the fields considered was studied in terms of timelines for the keywords, i.e. how the frequency of each keyword changed over time.

The proposed methodology can be summarized as follows:
\begin{enumerate}
\item Obtain the \emph{citation network} among the papers of the corresponding dataset.
\item Obtain the \emph{words}, corresponding to the $n$-grams present on both titles and abstracts of each paper. Here, we considered only unigrams and bigrams for the analysis. Also, we removed stop-words and the remaining words were lemmatized.
\item Apply a \emph{community detection algorithm} to the network, thus obtaining a partitioning of papers. Here, we opted to use a fast multilevel technique~\citep{blondel2008fast}.
\item Calculate the \emph{in-community frequencies}, $F^{in}_\alpha(w)$, for each word $w$ for all the communities, according to equation~\ref{eq:intraFrequency}.
\item Calculate the \emph{out-community frequencies}, $F^{out}_\alpha(w)$, according to equation~\ref{eq:outerFrequency}.
\item Calculate the \emph{importance index}, $I(w)$, of each word $w$ using equation~\ref{eq:keywordRanking}.
\item Sort the words according to the importance index and select an amount from the top. Here, we selected the first $50$ keywords to pair a similar amount of keywords provided by an expert.
\item Apply a hierarchical clustering method to the selected keywords, where the dissimilarity between two keywords corresponds to the average topological distance between papers presenting such words. This procedure results in the dendrogram of keywords.
\item The keywords can also be used to label the communities they belong to.
\item By using network visualization techniques, project the network to a 2D or 3D space and use the communities and the generated labels to obtain a scientific map~\citep{FRUCHTERMAN:1991fk,silva:2013,Bando:2013}.  In this work we employed the Fruchterman-Reingold algorithm and, for comparison purpose, we also use the VOSViewer~\citep{Eck:2010aa} visualization tool.
\end{enumerate}

It should be noted that the techniques employed in each step of our framework can be replaced by similar methods. For instance, one can use other visualization tools and techniques to construct science maps, or one can employ other community detection algorithms. While an extensive combination of techniques and parameters is still needed to uncover benefits and disadvantages of the framework, here we illustrate it by choosing only one set of methods and parameters. These correspond to the most traditional or simple methods required for each step.

\section{Results and Discussion}
We obtained two networks from the dataset, the CN network comprising $11,063$ papers with average degree $\langle k_{out}^{CN} \rangle \approx 8.5$, and the PC network encompassing $20,230$ papers and presenting $\langle k_{out}^{PC} \rangle \approx 6.6 $. Papers published from $1991$ to $2013$ were included in the networks. The structure of the CN network revealed $22$ communities yielding a modularity $q_{CN} \approx 0.53$, while $20$ communities were identified with modularity $q_{PC} \approx 0.65$ for the PC network.

\subsection{CN network analysis}
\begin{figure*}[!htb]
 \centering
  \includegraphics[width=18cm]{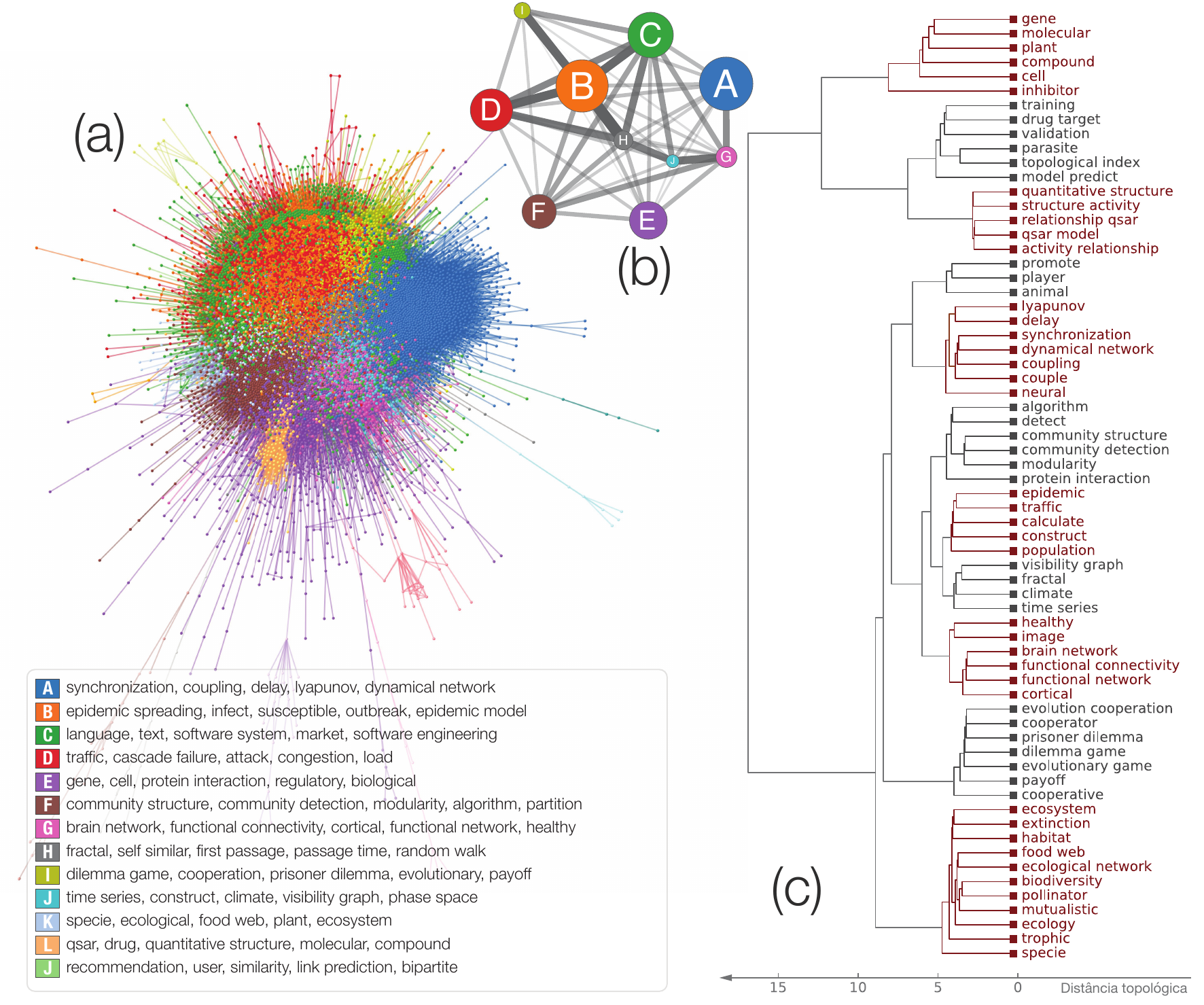}
   \caption{\label{fig:CN_Network} Projection of the CN network ({\bf a}) obtained by force-directed embedding with node colors representing the communities. The legends shows top keywords for each community ranked according to Eq.~\ref{eq:keywordRanking}. The relationships among communities obtained for the CN network are displayed in a coarse-grained diagram ({\bf b}). The diagram is obtained by collapsing each community in a single node with edges weighted by the fraction of original edges existing against all possible between two communities. Edges are represented by lines with thickness and intensity proportional to their weights. The top 50 keywords for the entire CN network are displayed in a dendrogram ({\bf c}) built with the hierarchical agglomerative clusterization method applied to the topological distance between the keywords.}
\end{figure*}

Fig.~\ref{fig:CN_Network}(a) displays the science map from the CN citation network, where the colors denote the communities associated with the top keywords according to the importance index of Eq.~\ref{eq:keywordRanking}. As expected by the high modularity, each module fills distinctive regions of the network topology. The only exception appears to be communities B and D that seem to share the same region, but this is an artifact of the 2D projection. A clear separation is confirmed in the 3D visualization (as shown in video~{\bf S1} in the supplementary material). It is interesting that most communities originate from a densely central region of the projection, as can be observed in the figure. This indicates that nodes at the central region are much more interdisciplinary.

The coarse-grained graph of the CN network is shown in Fig.~\ref{fig:CN_Network}(b), which features communities $B$, $C$ and $D$ strongly connected among themselves. Community $B$ (epidemic spreading dynamics) glues together many communities, being at the heart of the network alongside community $H$ (fractal, self-similar). This is probably because epidemic dynamics represented by community $B$ has a wide variety of applications in network science~\citep{Costa2011analyzing}. In spite of being the largest community, $A$ (synchronization and coupling) only connects strongly to $G$ (brain and cortical networks), highlighting the application of synchronization dynamics to modeling neuronal networks. Surprisingly, community $E$ (gene regulatory networks, protein interaction, etc) is the lesser connected among the communities. Besides, it presents no remarkable connection preference pattern, i.e. it is uniformly and weakly connected to other communities. This indicates that papers in this community still do not fully benefit from the tools and methodologies provided by network science.

The dendrogram obtained by clustering the top keywords, shown in Fig.~\ref{fig:CN_Network}(c), provided interesting insights. For instance, keywords from the field of ecological applications of complex networks associated with papers containing the words "ecosystem", "food web" and "biodiversity", are closely related among themselves. Although further investigations are needed to explain some counter intuitive exceptions such as the branch containing the keywords "promote", "player" and "animal", on the whole, the relationships between keywords are well described by the dendrogram and appears consistent with what should be expected from an expert in the area.

\begin{figure*}[!htb]
 \centering
\includegraphics[width=0.80\linewidth]{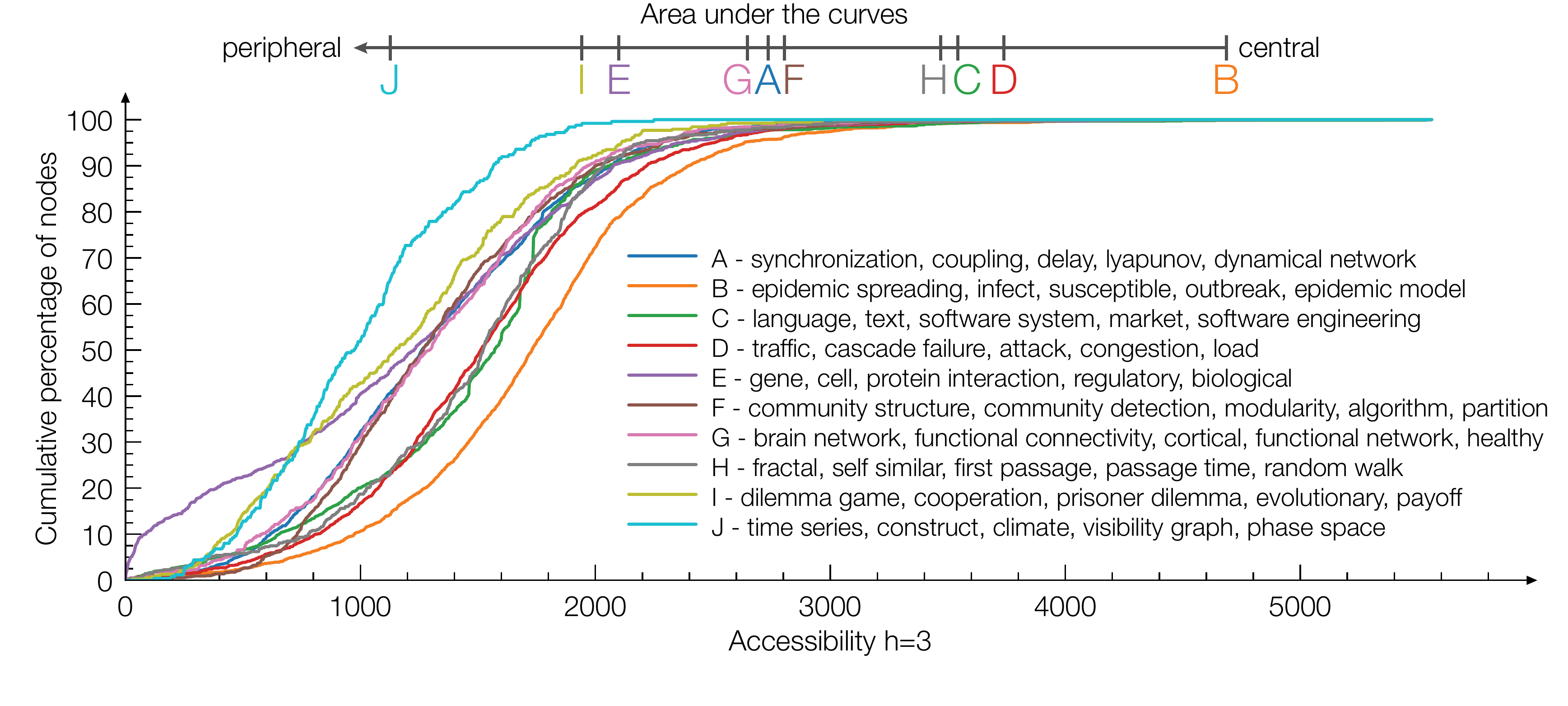}
 \caption{Curves of cumulative distribution of accessibility obtained for the CN network communities. The curves are presented in color according to the inset. On top of the figure the total area under the curves of each community is shown, which is related to the centrality or peripheral nature of its nodes.}
 \label{fig:communities-accessibility-CN}
\end{figure*}

The analysis was complemented using the \emph{accessibility} metric. The cumulative distribution of accessibility for $h=3$ taken over all nodes of the CN network is presented in Fig.~\ref{fig:communities-accessibility-CN}. We chose to calculate accessibility for level $h=3$ because node-centered measurements taken around the immediate neighborhood of a node (i.e. for $h=1$ or $h=2$) may depend on its degree~\citep{Costa:2006p278}. Also, because the networks are small-world, the measurement may suffer from border effects for large $h$. The data is grouped together by the community membership of nodes, hence each community has a different curve of cumulative accessibility distribution. With the data so presented it is easy to determine the percentage of nodes below or above a certain accessibility threshold. For instance, community $B$ possesses only roughly $10\%$ of nodes with accessibility $1000$ or lower.

We consider peripheral those communities containing many vertices with low accessibility. The area under the accessibility curves can be used to rank the communities according to their pertinence to the borders of the network. Communities covering a large area under the curves are at the boundaries of the network, as displayed on the top of Fig.~\ref{fig:communities-accessibility-CN}. Community $J$ (time series, climate and visibility graph) is the most peripheral, followed by $I$ (game, cooperation and prisoner dilemma) and $E$ (protein, gene and cell networks). In particular, community $E$ has about $20\%$ of papers with very low accessibility. Communities $G$ (brain and cortical networks), $A$ (synchronization and coupling) and $F$ (community structure and community detection) are close together and present average values of accessibility. The curves for $H$ (fractal, self similar and first passage), $C$ (language, text and software system) and $D$ (traffic, attack, cascade failure) also present similar patterns of accessibility among themselves and are much more at the core of the network than the aforementioned communities.

Corroborating the qualitative results from the analysis of the coarse grained graph, the most central community was $B$. The central core of the network is composed of communities related to techniques of network dynamics such as cascade failure, epidemic spreading dynamics and self-similarity techniques. On the borders are found more specific applications of networks such as cell networks cooperation and time series analysis.

\subsection{PC network analysis}
\begin{figure*}[!htb]
 \centering
  \includegraphics[width=18cm]{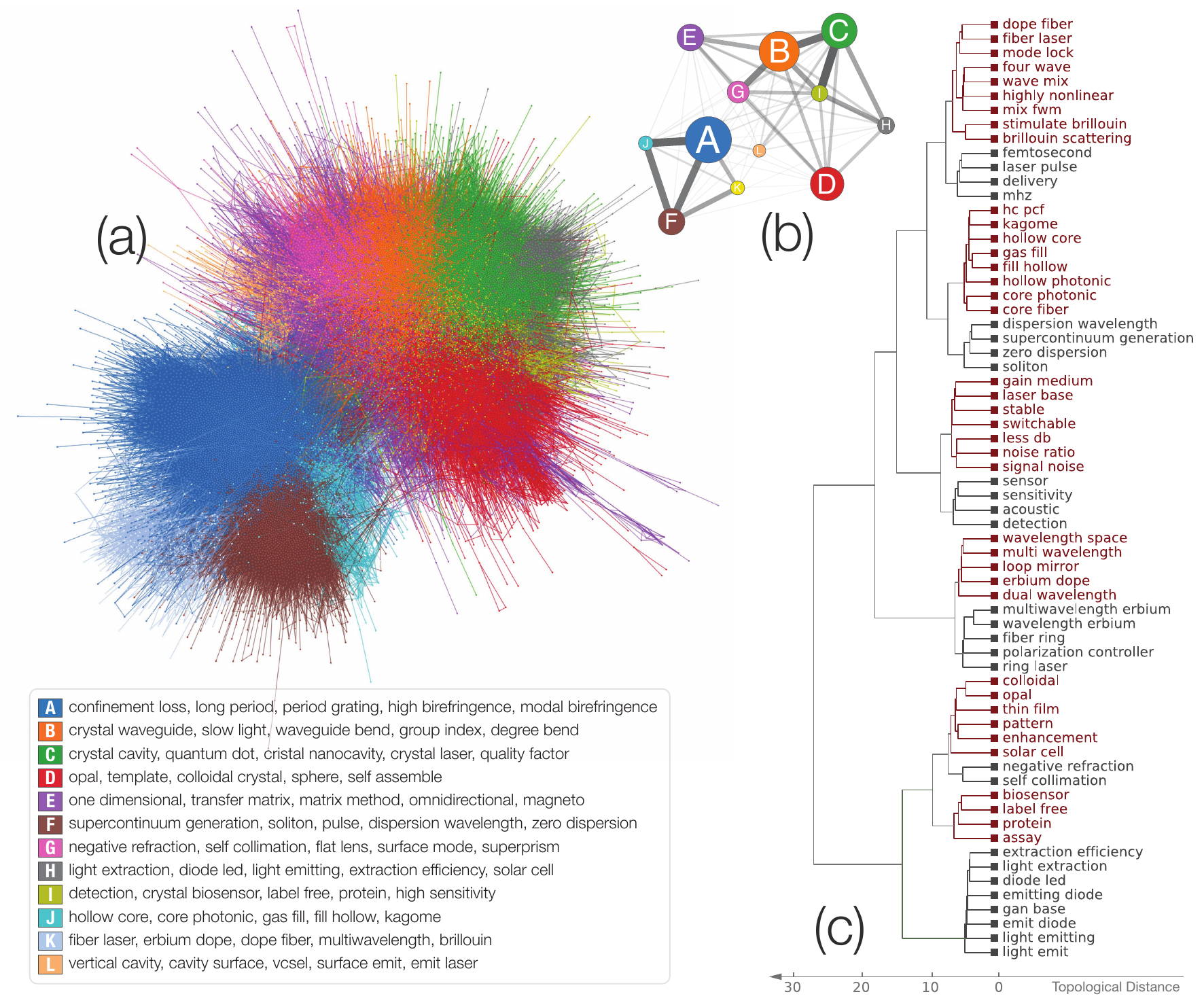}
   \caption{Projection ({\bf a}), coarse-grained diagram ({\bf b}) and keywords dendrogram ({\bf c}) of the PC network obtained in the same fashion as Fig.~\ref{fig:CN_Network}.}
 \label{fig:PC_Network}
\end{figure*}

The most striking feature of the science map represented by the PC network is its diploid nature, with two very distinct giant communities visualized in Figure~\ref{fig:PC_Network}(a). From the analysis of keywords associated with these giant communities it is readily noted that they refer to scientists from very distinct areas. The smaller giant community comprises papers from telecommunications, e.g. with keywords deriving from the photonic crystal fiber topic. Indeed, the keywords related to the communities from this giant community are (confinement loss, long period, high birefringence) for $A$, (supercontinuum generation, soliton) for $F$, (fiber laser, erbium dope, dope fiber) for $K$ and (porous silicon, silicon photonic, monitor) for $M$. The authors in this giant community are normally engineers exploiting fibers for telecommunications. The larger giant community is made of papers authored by experts in the development of the science of photonic crystals, mostly physicists. The interface between the two giant communities is quite thin, as shown in the figure, thus indicating little scientific interaction across the two enlarged communities.

The interface between the two giant communities is better visualized in the coarse-grained graph in Fig.~\ref{fig:PC_Network}(b), featuring connections from nodes in communities $E$ (one dimensional, transfer matrix, matrix method, omnidirectional), $G$ (negative refraction, self collimation), $I$ (detection, biosensor, label free) and especially $L$ (vertical cavity, cavity surface, vcsel, surface emit).
Also clear from the coarse-grained graph is the difficulty in establishing which communities are most central or peripheral owing to the diploid nature of the network.

\begin{figure*}[!htb]
 \centering
\includegraphics[width=0.80\linewidth]{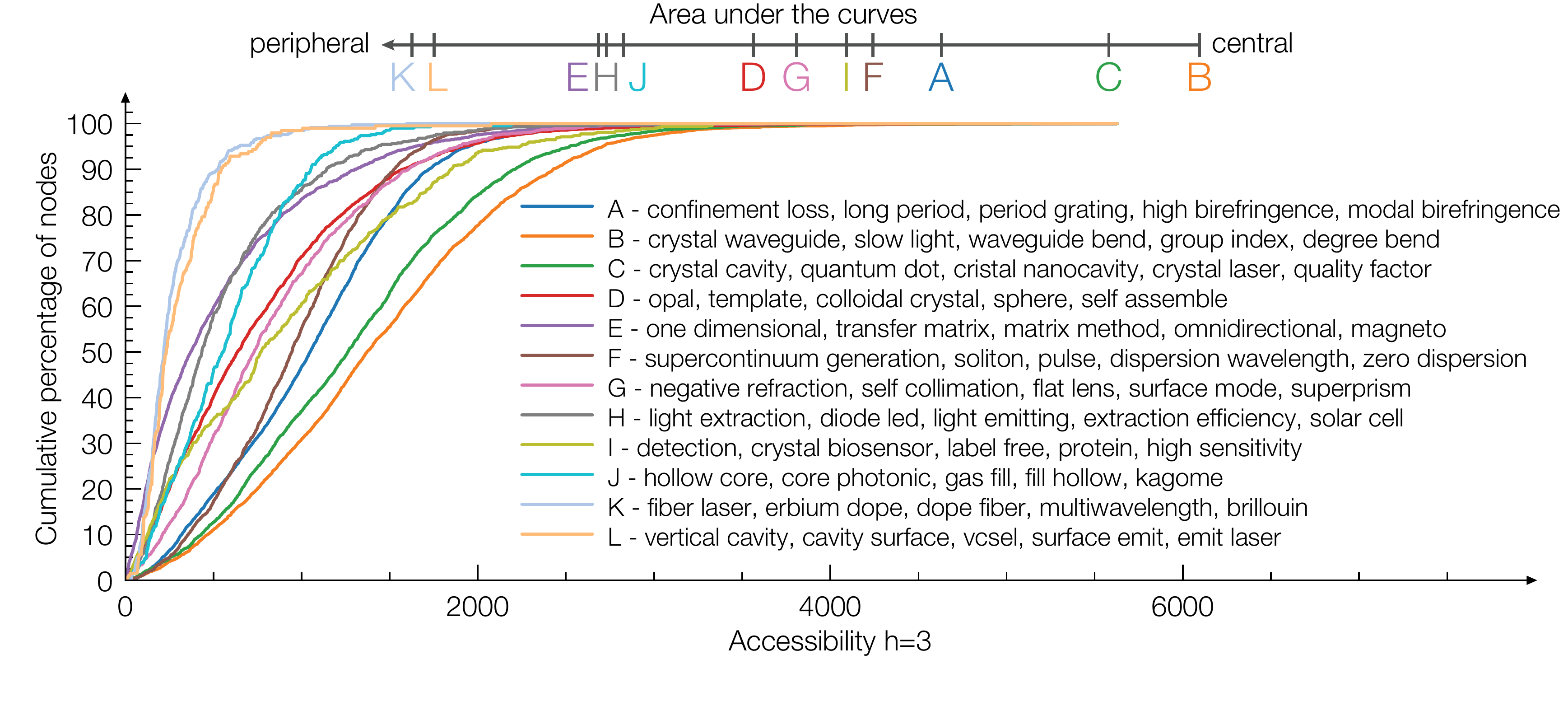}
 \caption{Cumulative accessibility distribution obtained for the PC network communities.}
 \label{fig:communities-accessibility-PC}
\end{figure*}

Here is a case where the accessibility metric is most useful. Because it is a local measurement, it avoids the pitfalls of other global centrality measurements when used to characterize networks presenting no well-defined border and central regions. When applied to the PC network, the analysis of cumulative accessibility in Figure~\ref{fig:communities-accessibility-PC} revealed that communities $K$ and $L$ are those most at the borders, followed by communities $E$, $H$ and $J$. Communities $C$ and $B$ are the most central in the network. Community $A$ can also be considered a central community on this smaller giant component. Analogously to what was observed for the CN network, general concepts of the PC field were found in the core of the system, such as papers of communities $B$ and $C$ comprising nodes having keywords "nanocavity", "quantum dot", "waveguide", "slow light", etc. On the other hand, more specific methodologies and applications are scattered on the borders of the network, such as in papers containing the keywords "fiber laser", "erbium dope", "vertical cavity", "transfer matrix", "one dimensional", etc.
The taxonomy reached by using the automated keywords for the PC network is consistent with expectation from experts, as indicated in the dendrogram of Fig.~\ref{fig:PC_Network}(c).
\begin{figure*}[!htb]
 \centering
  \includegraphics[width=18cm]{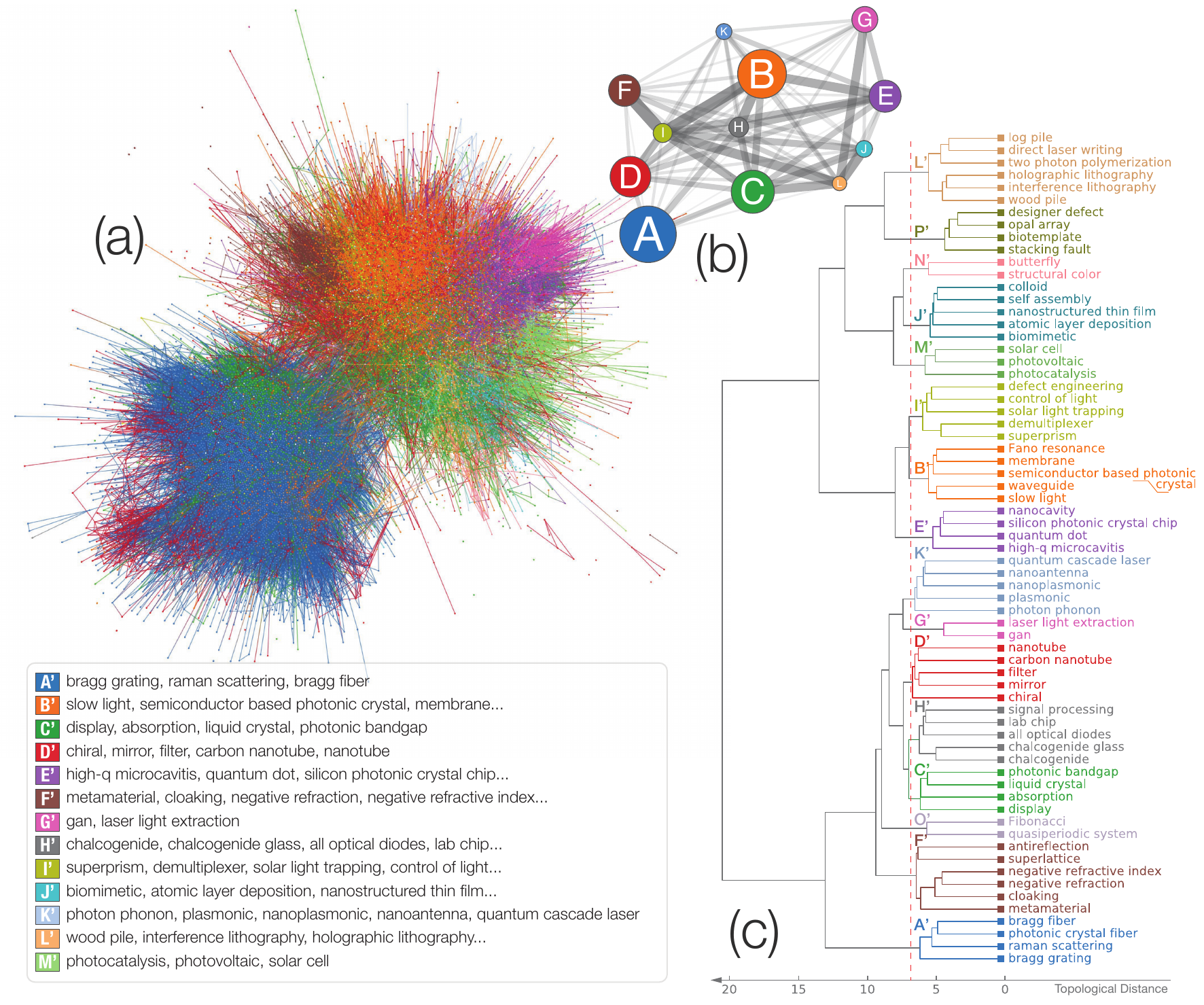}
   \caption{Projection ({\bf a}), coarse-grained diagram ({\bf b}) and dendrogram ({\bf c}) with the keywords provided by an expert for the PC network. Differently from Figs.~\ref{fig:CN_Network}, and \ref{fig:PC_Network}, the regions depicted by colors in ({\bf a}) correspond to the groups obtained after applying a threshold on the dendrogram as indicated by a dashed red line in ({\bf c}).}
 \label{fig:PC_Network-researcher}
\end{figure*}

We also used a list of $67$ keywords, containing up to $4$ words each, provided by one of the authors (MB), expert in the PC field. The dendrogram was constructed with the same approach as for the automated keywords in Fig.~\ref{fig:PC_Network-researcher}(c). It also provides valuable insights about the area, such as the fact that negative refraction index is closely related to \emph{metamaterials}, which in turn are key concepts for the technology that allows the development of an \emph{invisibility cloak}~\citep{Schurig2006Metamaterial,1367-2630-15-3-033037}. Another example concerns the keyword \emph{liquid crystal}, which appears, as expected, close to \emph{photonic bandgap}. A science map of the PC network was obtained using the expert´s keywords, where partitioning was reached by applying a threshold (as shown by the dashed line and group labels in Fig.~\ref{fig:PC_Network-researcher}(c)) to the dendrogram. The nodes were assigned to a community when their corresponding abstracts shared a large number of keywords that define a specific group. A comparison of Figs.~\ref{fig:PC_Network}(a) and~\ref{fig:PC_Network-researcher}(a) points to a narrower coverage of nodes for the keywords suggested by the expert for the small giant community associated with the telecommunications area. This was indeed expected because the expert (a physicist) has always worked with topics akin to the large giant community and had less familiarity with the use of photonic crystals in telecommunications. The coarse-grained network shown in Fig.~\ref{fig:PC_Network-researcher}(b) bears little resemblance to the one obtained from the community analysis of the network (\ref{fig:PC_Network}), with the groups of the former connecting strongly among themselves. However, a correspondence between some of the network communities and the groups of the expert´s keywords partitioning can be drawn by observing the communities sharing the same regions of the network (i.e. sharing a similar set of nodes). For instance, community $G$ (in Fig.~\ref{fig:PC_Network}(b)) shares the same region as the group $F'$ (in Fig.~\ref{fig:PC_Network-researcher}(b)), also displaying similar keywords, corresponding to subjects related to negative refraction and cloaking. In the same fashion, communities $C$ and $H$ share the same region of groups $E'$ and $G'$.

\begin{figure*}[!htb]
 \centering
  \includegraphics[width=16cm]{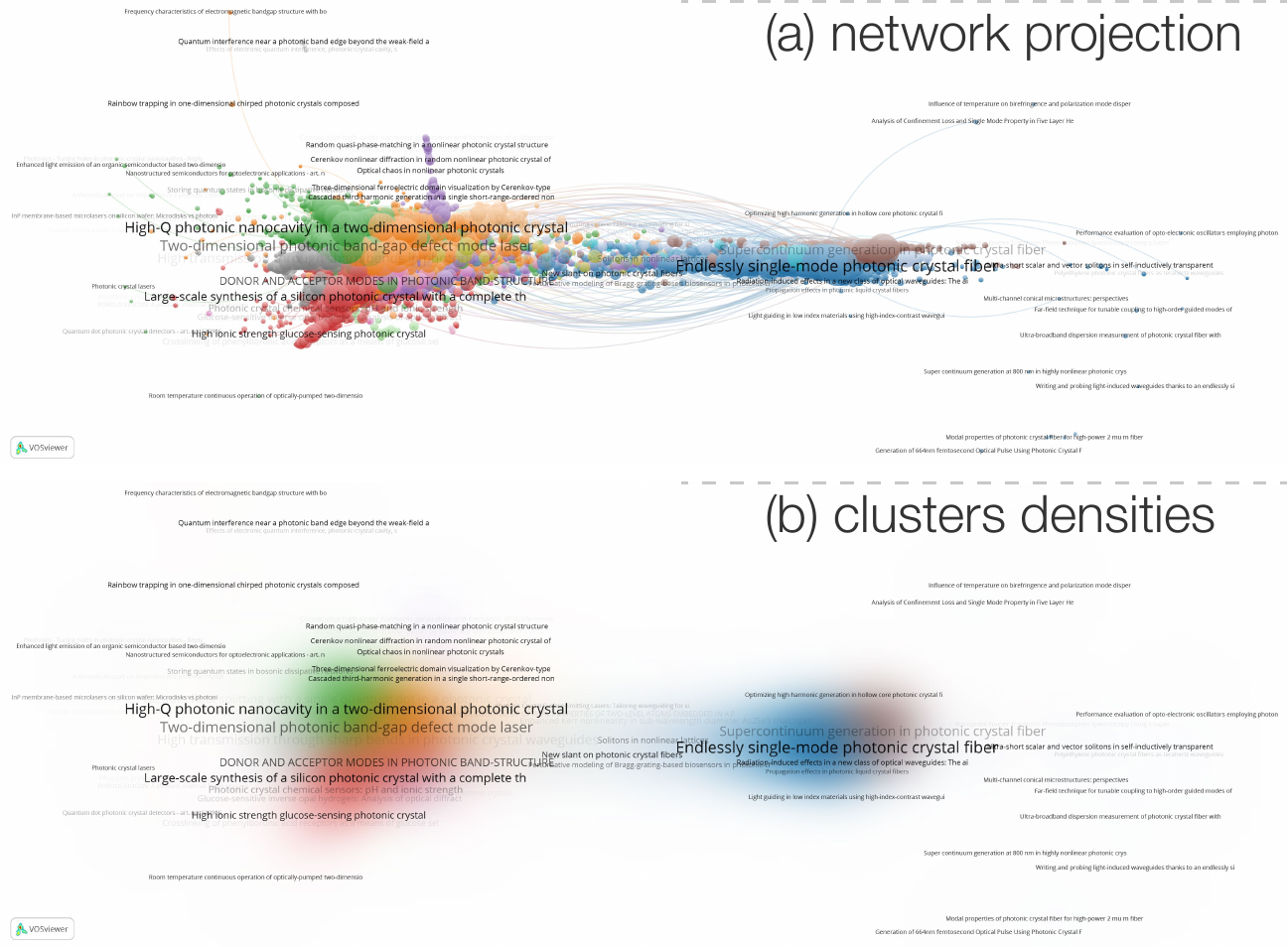}
   \caption{Visualization of the PC network using the VOSViewer visualization tool~\citep{Eck:2010aa}. The colors represent the same communities displayed in figure~\ref{fig:PC_Network}. Both the network ({\bf a}) and densities ({\bf b}) are shown.}
 \label{fig:PC_NetworkVosViewer}
\end{figure*}

To illustrate the possible replacement of methods in one of the steps of our framework, we also imported the network and labeled partitions into the VOSViewer tool~\citep{Eck:2010aa}. This visualization software has been used to construct scientific maps from network-based data encompassing a diverse range of disciplines and scientific fields. Figure~\ref{fig:PC_NetworkVosViewer} displays the projections attained by the software. The existence of the two major groups in the PC network is clearly more accentuated in the VOS Viewer visualization than by using the force-directed method, both in the positions (a) as well as in the density map (b). However, because of the anisotropic nature of the resulting map, some other aspects of the network structure cannot be observed clearly. For instance, it is difficult to tell how interconnected groups $A$ and $F$ are. In contrast, the isotropic nature of the maps obtained by the force-directed methodology reveals an informative interface between the two groups, which is reflected more clearly by the coarse-grained analysis. Nevertheless, the aims of the visualization techniques are different and may highlight distinct characteristics of the data. Perhaps the most useful approach is to use as many suitable visualization techniques as possible to draw better conclusions and attain deeper understanding of the datasets and of the analysis.


\begin{figure*}[!htbp]
 \centering
 \subfigure[][Complex Networks]{\label{fig:evolution-CN}\includegraphics[width=0.48\linewidth]{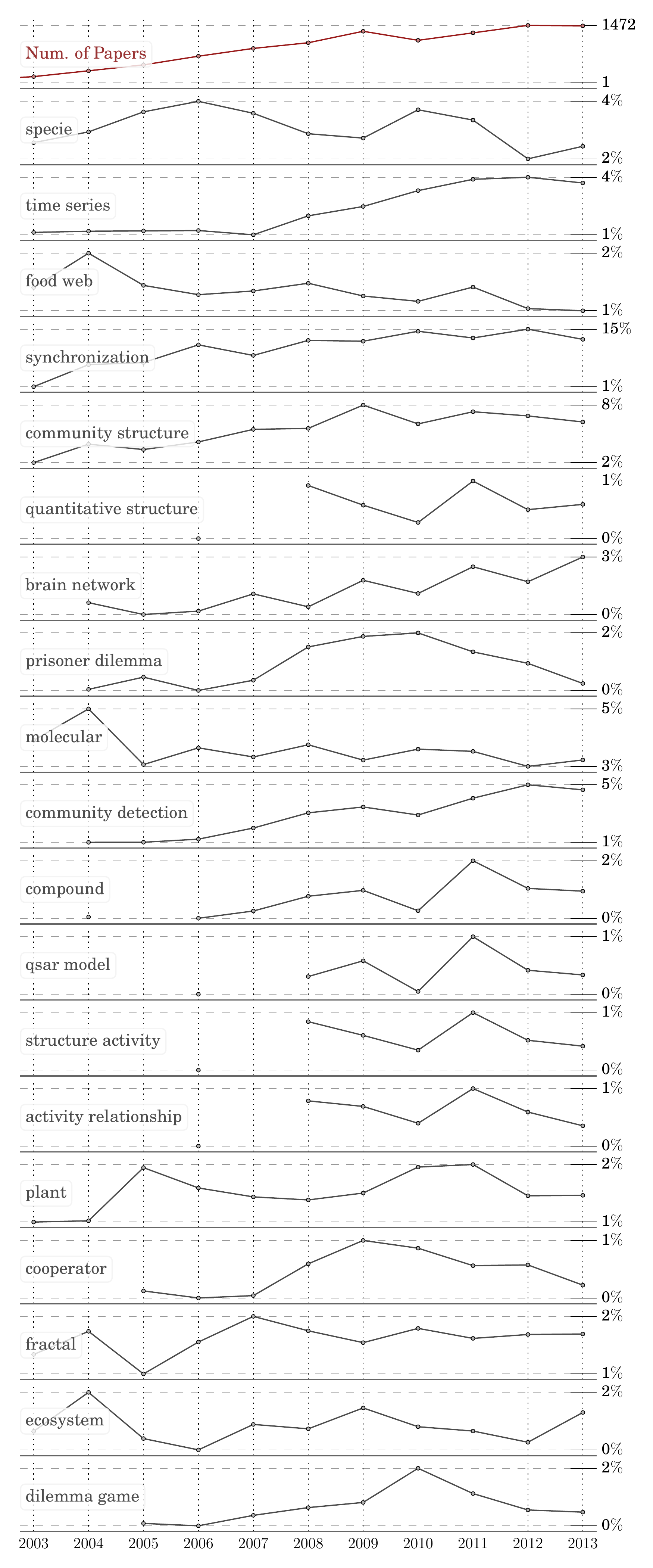}}
 ~
 \subfigure[][Photonic Crystals]{\label{fig:evolution-PC}\includegraphics[width=0.48\linewidth]{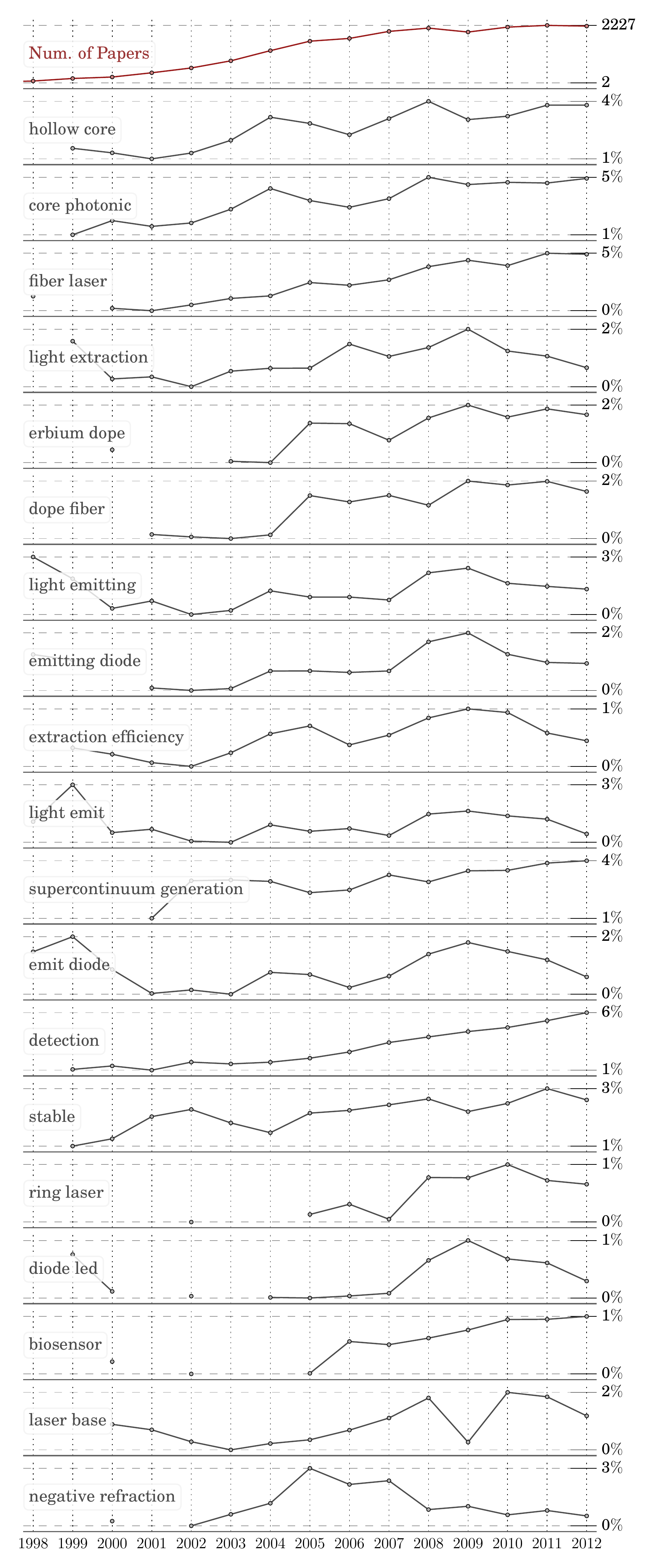}}
 \caption{Normalized frequency of occurrence for each keyword among papers published in the time period considered. The head graphics present the curves corresponding to the number of papers published in the corresponding years.}
 \label{fig:CN_PC_evolution}
\end{figure*}

The temporal evolution of the areas was examined by considering the timeline for the keywords. We counted the number of abstracts which contain the top keywords obtained from the ranking index in Eq.~\ref{eq:keywordRanking}. As the number of papers may greatly vary with the years, the frequencies were normalized by the total number of papers published in the same year. The resulting timelines are shown in Fig.~\ref{fig:CN_PC_evolution}. Because there are not many papers in the database for the years before $2003$ for the CN network, and $1998$ for the PC, only the subsequent years were considered.

The timelines confirm the extraordinary growth of both CN and PC areas (as shown on top of Fig.~\ref{fig:CN_PC_evolution}), but the growth rate decreased in the last few years. Several areas of CN have been growing: network applications to time series~\citep{lacasa2008fromtime,donner2010recurrence}, synchronization dynamics and analysis~\citep{arenas2008synchronization}, community detection~\citep{fortunato2010community}; while other subtopics are shrinking, such as food web and species networks~\citep{dunne2002foodweb}, cooperation dynamics~\citep{Yang2009diversity} and QSAR model~\citep{santana2008quantitative}. In PC field we can also observe distinct growth patterns. The subtopics hollow core photonic, fiber laser, erbium dope fiber, supercontinuum generation, detection, stable and biosensor are still growing on the network, while usage of terms light extraction efficiency, diode led and negative refraction are decreasing.

\section{Conclusion and Future Work}
The main goal of this paper was to introduce methods that could be used to automatically construct surveys on a given scientific field. We proposed a methodology to simultaneously analyze contextual information (in terms of papers abstracts) and citation networks, and this was applied to two fields: Complex Networks and Photonic Crystals. Upon identifying communities, it was possible to generate a taxonomy for these fields.

Several patterns could be inferred from the results. For complex networks, for instance, border communities were found to be related to regulatory and protein-protein interaction networks, in addition to subtopics related to climate, time series and visibility graphs. The interpretation is that these subtopics are not fully explored, at the moment, by the many complex networks analysis methods.

The PC network was peculiar in featuring two giant communities, each of which could be identified by analyzing the keywords. As expected, we found that one giant community comprises telecommunication engineers who use photonic crystal fibers in their applications, while the other, larger community is composed mainly of physicists. Surprisingly, not much interaction exists between the two communities, and this piece of information may be valuable to foster collaboration in the future.

The approach proposed here to construct the taxonomy for a survey differs significantly from what exists in the literature. Instead of using only similarities between terms of each abstract, here a citation network was used to provide both the distance among terms and the clustering (derived from the community structure). In addition, a simple text analytics technique was employed to provide the salience of terms according to the obtained community structure.

Here, we did not compare our results to those obtained from traditional text analytics techniques, particularly because the methods address two different classes of problems. Our approach takes into consideration how, in practice, researchers refer to other works in their fields, which may differ significantly from the similarity of terms obtained using only the textual content. The discrepancy between cited works and their contextual similarity has been a recent topic of study, with an in-depth analysis ~\citep{Amancio:2012aa,Ciotti:2016aa}. We understand that the organization of the scientific community, i.e., the citation patterns among researchers and papers, must play an important role for constructing a survey in a science field. In this context, our approach is more suitable for this task than methods based solely on text similarity.

We can still compare the technical limitations of the approach presented here and of those based on text analytics. For instance, one of the main disadvantages of topic analysis is the high computational cost involved in estimating the Markov model, which requires several iterations of Gibbs Sampling. This kind of analysis precludes the study of bigrams and higher order $n$-grams, while our approach can be extended to account for $n$-grams. In addition, the limitations of such analysis are not yet completely understood~\citep{icml2014c1_tang14}. Other methods such as those based on supervised learning need the input of annotated corpus or sets of golden summaries, which are not commonly available in scientific datasets. We however should point out that our approach is strongly dependent on the chosen network structure. If a co-authorship network among papers was used, instead of the citation network, the results should be interpreted in a different direction and could not be used, for instance, to construct a survey. As for topic analysis, an extensive study of the limitations of our approach is still needed to identify its strengths and disadvantages.

Several extensions of the approach we presented can be performed in future works. For simplicity, we did not consider the direction of the citation networks or the strongly asymmetric nature of the networks. These features could play an important role in the understanding of how distinct fields interact among themselves by citations.

In our methodology we did not take into consideration the importance and redundancy of papers. These limitations may be surpassed by using topological characterization at the level of papers. Future research should also address the problem of quantifying the interdisciplinarity. It is hoped that the approach inherent in the methods we introduced can be applied to build new tools and assist researchers in understanding their own or new specialty areas.

\section*{Acknowledgments}
L. da F. Costa thanks CNPq (grant no. 307333/2013-2) and FAPESP-MCT/CNPq/PRONEX (grant no. 11/50761-2) for support. F. N. Silva acknowledges CAPES and FAPESP (grant no. 15/08003-4). D. R. Amancio thanks FAPESP (grant no. 14/20830-0). O. N. Oliveira Jr acknowledges FAPESP and CNPq. M. Bardosova acknowledges Science Foundation Ireland for support.


\bibliography{paper}

\end{document}